\documentclass[review,12pt]{elsarticle}
\usepackage{lineno,hyperref}
\modulolinenumbers[5]
\usepackage{url}
\journal{Journal of \LaTeX\ Templates}
\usepackage{amsmath}
\usepackage{graphicx}
\usepackage{mathtools}

\title{Absorption of acoustic phonons in  Fluorinated Carbon 
Nanotubes with non-parabolic, double periodic band.}

\author[els]{D. Sekyi-Arthur\corref{cor1}}
\author[els]{S. Y. Mensah}
\author[rvt]{ N. G. Mensah}
\author[]{\\}
\author[els]{ K. A. Dompreh}
\author[els]{R. Edziah}
\address[els]{Department of Physics, College of Agriculture and Natural Sciences, U.C.C, Ghana.}
\address[rvt]{Department of Mathematics, College of Agriculture and Natural Sciences, U.C.C, Ghana}
\cortext[cor1]{Corresponding author\\ Email: daniel.sekyi-arthur@stu.ucc.edu.gh,rasdean3@gmail.com\\
This work is licensed under the Creative Commons Attribution International\\ License (CC BY).
\url{http://creativecommons.org/licenses/by/4.0/} }
\ead[url]{rasdean3@gmail.com}

\date{}

\begin{document}
\begin{abstract}
We studied theoretically the absorption of acoustic phonons in the hypersound regime in Fluorine 
modified Carbon Nanotube (F-CNT) $\Gamma_q^{F-CNT}$ and compared it to that 
of undoped Single Walled Nanotube (SWNT) $\Gamma_q^{SWNT}$. 
Per the numerical analysis, the F-CNT showed less absorption to that of SWNT
thus $\vert\Gamma_q^{F-CNT}\vert  < \vert\Gamma_q^{SWNT}\vert $. 
This is due to the fact that  Fluorine is highly electronegative 
and weakens the walls of the SWNT. Thus, the $\pi$-electrons associated with 
to the Fluorine which causes less free charge carriers to interact with 
the phonons and hence changing the metallic properties of the SWNT to 
semiconductor by the doping process. 
From the graphs obtained, the ratio of  hypersound absorption in SWNT 
to F-CNT at $T = 45K$ is $\frac{\Gamma_{(SWNT)}}{\Gamma_{(F-CNT)}}\approx 29$ 
whilst at $T = 55K$, is $\frac{\Gamma_{(SWNT)}}{\Gamma_{(F-CNT)}}\approx 9$ and 
at $T = 65K$, is $\frac{\Gamma_{(SWNT)}}{\Gamma_{(F-CNT)}}\approx 2$.
Clearly, the ratio decreases as the temperature increases.

\end{abstract}

\maketitle
\section{Introduction}
Recently, acoustic effects in bulk and low dimensional 
materials  have attracted lots of attention. This is due to the need to 
find coherent acoustic phonons for scientific applications  as against 
the use of direct current~\cite{1}. 
Materials such as homogenous semiconductors, Superlattices (SL), Graphene and Carbon 
Nanotubes (CNT) are good candidates for such studies due to their properties  
such as the high scattering mechanism, the high-bias mean-free path ($l$) and their 
sizes which enable strong electron-phonon interaction to occur in them 
resulting  in acoustic phonon scattering. 
Acoustic waves through these materials are characterized 
by a set of elementary resonance excitations and dynamic non-linearity which 
normally lead to  an absorption (or  amplification),    
Acoustoelectric Effect (AE)~\cite{2} and  Acoustomagnetoelectric Effect (AME)~\cite{3,4}.
The concept of acoustic wave amplification in bulk materials
was predicted by Tolpygo and Uriskii(1956)~\cite{5}, and in N-Ge by Pomerantz~\cite{6}. 
In SL, Mensah et. al. ~\cite{7} studied hypersound absorption(amplification) 
and established its use as a phonon filter and in~\cite{8} predicted the 
use of the SL as hypersound generator which was confirmed in~\cite{1}. 
In Graphene, Nunes et. al ~\cite{9} treated theoretically hypersound amplification but 
Dompreh et.al.~\cite{10} further proved that absorption also occurs in the material. Experimentally,  
Miseikis et. al.~\cite{11}, Bhandu and Nash~\cite{12} studied acoustoelectric effect in Graphene.\\ 

Carbon Nanotubes (CNT) on the other hand,  are cylindrical hollow rod of graphene sheets  whose 
electronic structures are determined by the localized $\pi$-electrons in the $sp^2$-
hybridized bonds. Absorption (Amplification) of hypersound in undoped CNT has been carried out 
theoretically by Dompreh et. al.~\cite{13,14} and experimentally by~\cite{15,16}. Other forms 
of research such as hot-electron effect~\cite{17}, thermopower in CNT~\cite{18} have been carried out. 
Flourine-modified CNT
(F-CNT) is off-late attracting a lot of scientific interest. This is attained 
by doping  the CNT with Fluorine thus  forming double periodic band CNT 
changing from metallic to semiconductor.
As per the studies conducted by Jeon et. al. ~\cite{19}, 
absorption in F-CNT is less than that of SWNT but
no studies have been done on
the absorption of F-CNT in the hypersound regime.
In this paper, the study of absorption 
of acoustic phonons in  metallic SWNT and F-CNT are theoretically studied.
Here, the acoustic wave  considered 
has wavelength $\lambda = 2\pi/q$, smaller than the mean-free path of the 
CNT  and then treated as a packet of coherent phonons (monochromatic
phonons) having a $\delta$-function distribution as
\begin{eqnarray}
N(k)=\frac{(2\pi)^{3}}{\hbar\omega_{q}V_{s}}\phi\delta(k-q)
\end{eqnarray}
where $k$ is the phonon wavevector, $\hbar$ is the Planck's constant 
divided by $2\pi$, and $\phi$ is the sound flux density, and $\omega_{q}$ 
and $V_{s}$ are respectively the frequency and the group velocity of 
sound wave with wavevector $q$. It is assumed that the sound wave is
propagated along the $z$-axis of the CNT.\\

This paper is organized as follows: In section 2, the absorption 
coefficient for F-CNT and SWNT   are calculated.
In section 3, the final equations are analyzed numerically 
and presented  graphically. Section 4 presents the conclusion of the study.

\section{Theory}
Proceeding as in \cite{20}, the acoustic phonon absorption coefficient is given as 
\begin{eqnarray}
\Gamma_q=\frac{2\pi\phi}{\omega_{q}V_{s}}.\frac{\Lambda^{2}q^{2}}{2\sigma\omega_{q}}\int[f(\varepsilon(p_z+\hbar q))-f(\varepsilon(p_z))]\delta(\varepsilon_{p_{z}+q}-\varepsilon_{p_{z}}-\hbar\omega_{q})dp_z
\end{eqnarray}
where $f(p_z)$ is the unperturbed  Boltzmann  distribution function,
$p_z$ is the phonon momentum, $\varepsilon(p)$ is the energy dispersion, $\Lambda$ is the deformation
potential  and $\sigma$ is the density of the CNT. 
The distribution function is given as 
\begin{eqnarray}
f(p_z) = C\exp[-\beta(\varepsilon(p_z)-\mu)]
\end{eqnarray}
where $\mu$ is the chemical potential of the system, $\beta = (k_B T)^{-1} $,
$k_B$ is the Boltzmann constant, $T$ is the absolute temperature and $C$ is 
the normalization constant to be determined from the normalization condition $\int f(p)dp = m$ as
\begin{eqnarray}
C = \frac{ma^{2}}{4\pi^{2}I_{o}(2\gamma_{o}\beta)I_{o}(6\gamma_{o}\beta)}\exp[\beta(\alpha_{\pi}-\mu)]
\end{eqnarray}
$m$ is the surface concentration of charge carriers and $I_o$ is the 
modified Bessel function. For a chemically modified F-CNT, where the Fluorine atoms form 
a one-dimensional chain, the energy dispersion can be deduced by 
using the Huckel matrix method where translational symmetry is 
accounted for   as ~\cite{21}
\begin{equation}
\varepsilon = \alpha_{\pi} +  \Xi_n\gamma_0 \cos^{2n-1} (a p_z) 
\end{equation}
Here $a = \sqrt{3}a_{c-c}/(2\hbar)$, $\Xi$ is a constant, $n$ is an integer and $\alpha_{\pi}$ is the minimum energy of the 
$\pi-$electrons within the first Brillouin zone.
For $n = 2$, the energy dispersion for F-CNT at the Fermi surface 
at the edge of the Brillouin zone is  
\begin{eqnarray}
\varepsilon(p_z)=\alpha_{\pi} + 8\gamma_{o}\cos^{3}(ap_{z})
\end{eqnarray}
From energy conservation principle, the momentum ($p_z$)  can be deduced 
from the delta function part of Eqn.($2$) as 
\begin{eqnarray}
p_{z}=-\frac{\hbar q}{2}+\frac{1}{4a}\sin^{-1}\left(\frac{\omega_{q}}{12\gamma_{o}a q}\right)
\end{eqnarray}
By substituting $p_z$ into the distribution function in Eqn.($2$), and
after some cumbersome calculations yields 
\begin{multline}
\Gamma_q^{F-CNT}=\Gamma_o
\left[\sinh\left\{2\gamma_{o}\beta\sin\left(\frac{3}{2}a\hbar q\right)\sin{A}+6\gamma_{o}\beta\sin\left(\frac{a}{2}\hbar q\right)\sin{B}\right\}\right.\\
\times\left.\cosh\left\{2\gamma_{o}\beta\cos\left(\frac{3}{2}a\hbar q\right)\cos{A}+6\gamma_{o}\beta\cos\left(\frac{a}{2}\hbar q\right)\cos{B}\right\}\right]
\end{multline}
where
\begin{eqnarray*}
\Gamma_o=\frac{ma^2}{48\pi^{2}I_{o}(2\gamma_{o}\beta)I_{o}(6\gamma_{o}\beta)}\frac{\pi\phi\Lambda^{2}q}{\omega_{q}^{2}\sigma V_{s}\gamma_{o}\hbar }\frac{\Theta(1-\alpha^{2})}{\sqrt{1-\alpha^{2}}}
\end{eqnarray*}
where $\Theta$ is the Heaviside step function
\begin{eqnarray*}
A=\frac{3}{4}\sin^{-1}\left(\frac{\omega_{q}}{12\gamma_{o}a q}\right)&
B=\frac{1}{4}\sin^{-1}\left(\frac{\omega_{q}}{12\gamma_{o}a q}\right)&
\alpha=\frac{\omega_q}{12\gamma_{o}aq}
\end{eqnarray*}
To compare the result with an undoped SWNT, we follow the same procedure 
as that of F-CNT.  Using the tight-binding energy dispersion of the 
$p_z$ orbital which is given  as:
\begin{eqnarray}
\varepsilon(p_z) = \pm \gamma_{o}\sqrt{1+4\cos\left(\frac{\nu\pi}{n}\right)\cos\left(\frac{p_z\sqrt{3}a_{c-c}}{2\hbar}\right)+4\cos^2\left(\frac{p_z\sqrt{3}a_{c-c}}{2\hbar}\right)}
\end{eqnarray}
where $\gamma_o = 2.6\mathrm{eV}$ is the hopping integral parameter, $a_{c-c} = 0.142\mathrm{nm}$ 
is the C-C bonding distance, and ($+$) and ($-$) signs are respectively the conduction and valence 
band. When $\nu = 0$, the conduction and valence bands cross each other near the Fermi points, 
$p_F = \pm 2\pi\hbar/3\sqrt{3} a_{c-c}$ giving the metallic nature to the armchair tube. Putting 
$\nu = 0$, and making the substitution, $p_z = p_z+3p_{o}/2\hbar$ in Eqn.($9$) gives
\begin{eqnarray}
\varepsilon(p_z) = \pm \gamma_{o}\left(1-2\cos\left(\frac{p_{z}\sqrt{3}a_{c-c}}{2\hbar}\right)\right)
\end{eqnarray}
where $p_{o} = 2p_{F} = 4\hbar\pi/3\sqrt{3}a_{c-c}\approx 1.7\times 10^{10}\mathrm{m^{-1}}$ see~\cite{22}. 
Eqn.($10$)  is equivalent to the energy dispersion in Eqn.($5$) 
when  $n = 1$, which is   
\begin{eqnarray}
\varepsilon(p_z)=\alpha_{\pi} + \Xi\gamma_{o}\cos(ap_{z})
\end{eqnarray}
Using Eqn.($10$), the absorption in SWNT is calculated as 
\begin{multline}
\Gamma_q^{SWNT} = \frac{\pi^2\Lambda^{2}q^{2}\phi^2}{4\gamma_{o}^2\omega_{q}^2V_{s}\sigma a\sin(\frac{a\hbar q}{2})}\frac{na^{2}}{I_{o}(2\gamma_{o}\beta)}\\
\times\sinh\left\{\beta\hbar\omega_q\right\}\cosh\left\{4\gamma_{o}\beta\sqrt{1-\alpha^{2}}\cos\left(\frac{a\hbar q}{2}\right)\right\}\frac{\Theta(1-\alpha^{2})}{\sqrt{1-\alpha^{2}}}
\end{multline}
where
\begin{eqnarray*}
\alpha=\frac{\hbar\omega_q}{4\gamma_{o}\sin(\frac{a\hbar q}{2})}
\end{eqnarray*}

\section{Results and Discussion}
The general expressions for the hypersound absorption in F-CNT ($\Gamma_q^{F-CNT}$)
and in SWNT ($\Gamma_q^{SWNT}$) are presented in Eqn.($8$) and Eqn.($12$)
respectively. In both equations, the absorptions are dependent on the 
frequency ($\omega_q$), the acoustic wavenumber ($q$), and temperature ($T$)
as well as other parameters such as the inter-atomic distances, the velocity
of sound ($V_s$) and the deformation potential ($\Lambda$). 
In both expressions (see Eqn.($8$) and Eqn.($12$)) a transparency window
is observed: for F-CNT is $\omega_q >> 12\gamma_o a q$;
and for SWNT is $\omega_q >> \gamma_o \sin(\frac{1}{2}a\hbar q)/\hbar$.
These are the  consequence of conservation laws. The Equations 
($8$) and ($13$), are analyzed numerically  with the following parameters 
used:  $\Lambda=9\mathrm{eV}$, 
$q = 10^{5}\mathrm{m^{-1}}$, $\omega_{q} = 10^{12}\mathrm{s^{-1}}$, $V_{s} = 5\times10^{3}\mathrm{m/s}$,
$\phi = 10^{4}\mathrm{Wb/m^{2}}$, and $T = 45\mathrm{K}$.
 The results are graphically plotted (see Fig. $1$, $2$, and $3$). 
\begin{figure}[h!]
\includegraphics[width = 8.0cm]{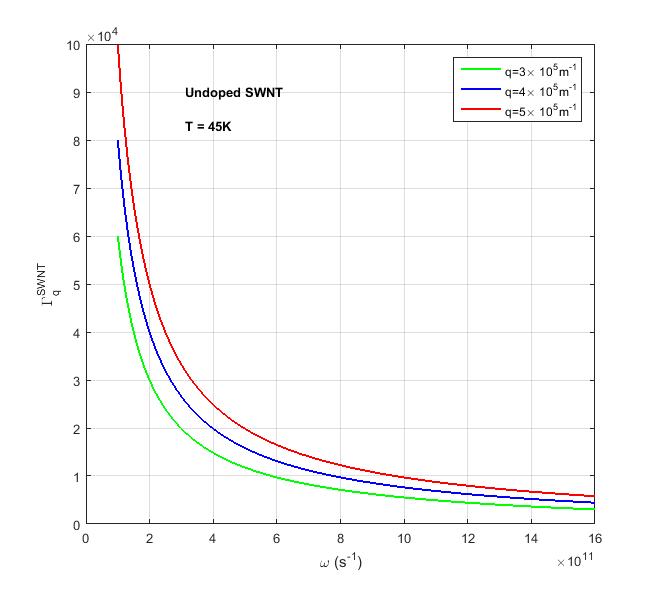}
\includegraphics[width = 8.0cm]{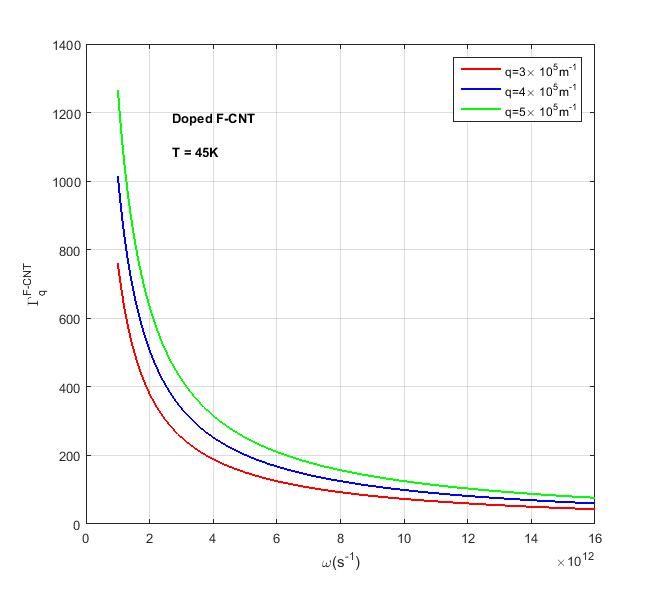}
	 \caption{Dependence of $\Gamma$  on $\omega_{q}$ (left) 
for an undoped SWNT,  (right) for a F-CNT by varying $q$ at $T = 45K$}
\end{figure}

Figure $1$ shows the dependence of the sound absorption coefficient 
on the the frequency ($\omega_q$) for varying $q$. 
In both graphs, the absorption is initially high but falls off sharply 
and then changes slowly at high values of $\omega_q$. Increasing the values 
of $q$ correspondingly increased the obtained graph in both  doped F-CNT and 
undoped SWNT though the magnitude of absorption obtained in SWNT exceeds 
that of F-CNT, that is, $\vert\Gamma_q^{SWNT}\vert > \vert\Gamma_q^{F-CNT}\vert$. 
This is in accordance with the work of  Jeon et. al. ~\cite{19}.   
 \begin{figure}[h!]
		\includegraphics[width = 8.0cm]{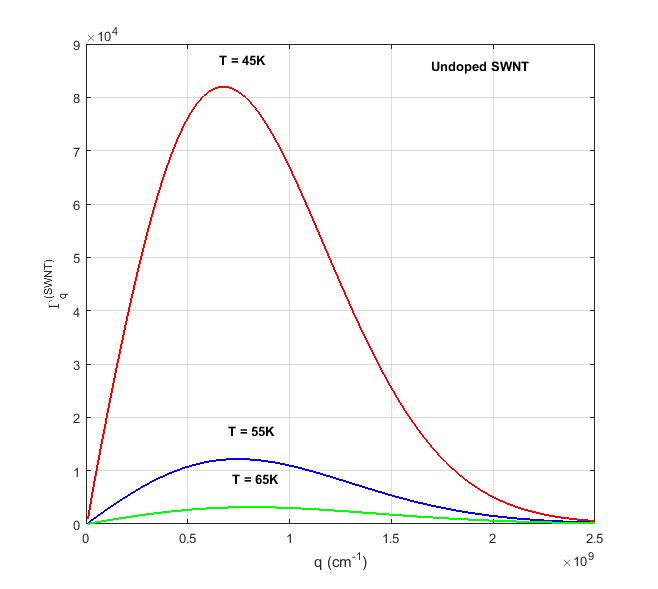}
\includegraphics[width = 8.0cm]{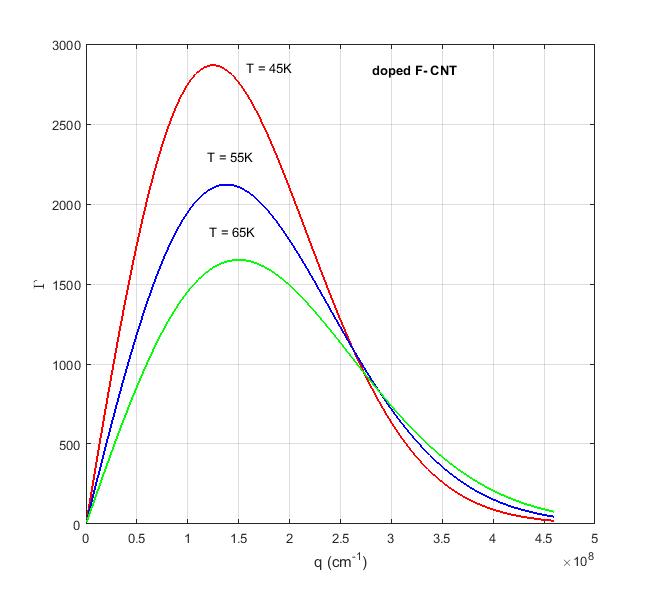}
	 \caption{Dependence of $\Gamma$  on $q$ (left) for 
undoped SWNT and (right) for doped F-CNT at $T = 45, 55, 65K$}
\end{figure}
In Figure $2$, the graph increases to a maximum point then drops off. It 
then changes again slowly at high $q$ 
for both undoped SWNT and doped F-CNT. By increasing the
temperature, the the amplitude of the graphs reduces. For $T = 45K$, the maximum 
absorption in  $\Gamma_q^{SWNT} = 8.2\times 10^4$ whilst that of
$\Gamma_q^{F-CNT} = 2867$ which gives the ratio of the absorption 
$\frac{\Gamma_{(SWNT)}}{\Gamma_{(F-CNT)}}\approx 29$, whilst at $T = 55K$, 
$\frac{\Gamma_{(SWNT)}}{\Gamma_{(F-CNT)}}\approx 9$ and at $T = 65K$, we had 
$\frac{\Gamma_{(SWNT)}}{\Gamma_{(F-CNT)}}\approx 2$. Clearly, we noticed that the 
ratio decreases with an increase in temperature. 
To aid a better understanding of the comparison between the absorption obtained in both SWNT and F-CNT,
a semilog plot is presented in Figure $3$ which clearly shows that the undoped
SWNT absorbs more than the doped F-CNT.
 \begin{figure}[h!]
		\includegraphics[width = 8.2cm]{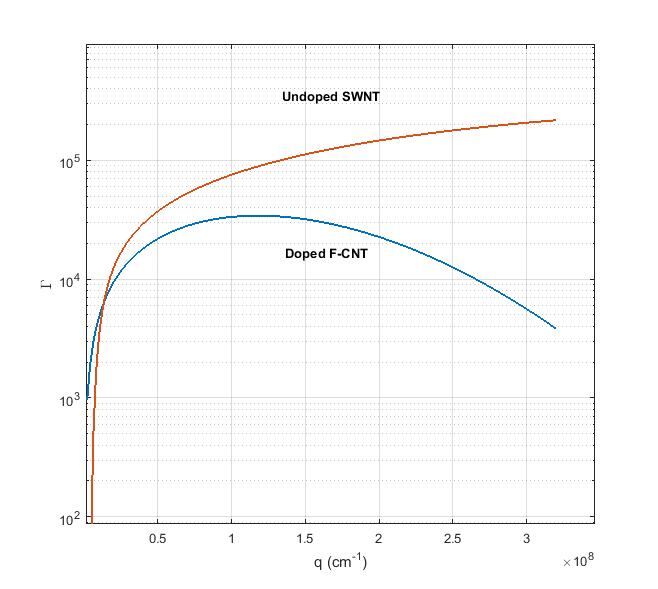}
\includegraphics[width = 8.2cm]{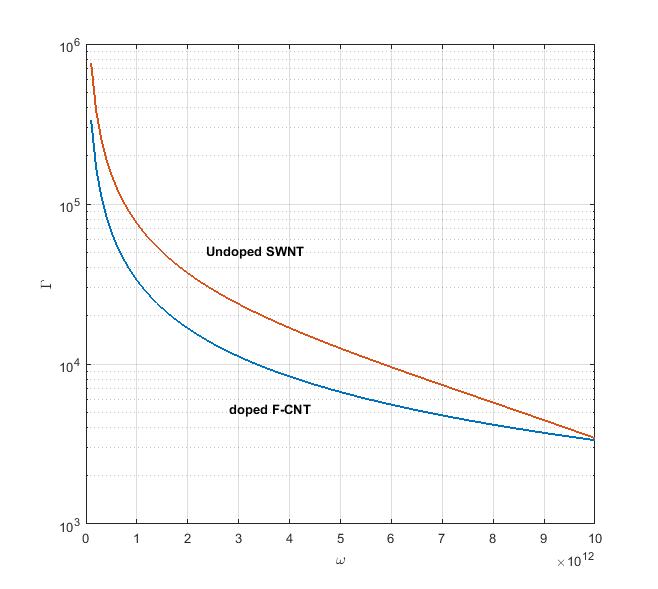}
	 \caption{Semilog plot  of $\Gamma$ dependence on $q$ and $\omega_q$ for doped 
F-CNT and undoped SWNT}
\end{figure}
This can be attributed to the fact that the presence of F-CNT atoms leads to chemical activation of a passive surface CNT by adding
additional electronic band structure and altering the carbon $\pi$-bonds around
the Fermi level in a non-linear manner  thus forming a band structure of
width two periods~\cite{21}. 
As Flourine is highly electronegative it thus  weakens the walls of the CNT as it 
approaches it. The $\pi$-electrons attached to the Flourine which causes less free charge
carriers to interact with the phonons. Current researches have predicted 
$sp^2$  bonding charge change to $sp^3$ by F-functionlization~\cite{23,24,25}.
This bonding charge change would reduce the density of free carriers, 
consequently leading to the magnitude reduction of the absorption ~\cite{21}.

\section{Conclusion}
The absorption of hypersound in F-CNT and SWNT 
was theoretically calculated in the regime $ql >> 1$. From the numerical analysis,  the graphs of 
the absorptions in F-CNT and SWNT are plotted and compared.   The Flourine functionalization   
affects  the absorption properties of F-CNT. The SWNT absorbs more than the 
F-CNT as was observed by Jeon et. al.

\end{document}